\documentclass[pre,twocolumn,
superscriptaddress,showpacs,showkeys,amsmath,amssymb]{revtex4}
\usepackage{graphicx}
\usepackage{dcolumn}
\usepackage{bm}

\begin{document}

\title{Single-Cycle High-Intensity Electromagnetic Pulse Generation in the \\
Interaction of a Plasma Wakefield with Nonlinear Coherent Structures}

\author{S. S. Bulanov}
\affiliation{University of Michigan, Ann Arbor, USA}
\affiliation{Institute of Theoretical and Experimental Physics, Moscow, Russia}

\author{T. Zh. Esirkepov}
\affiliation{Moscow Institute of Physics and Technology, Dolgoprudnyi,
Moscow Region, Russia}
\affiliation{Kansai Photon Science Institute, JAEA, Kizu, Kyoto, Japan}

\author{F. F. Kamenets}
\affiliation{Moscow Institute of Physics and Technology, Dolgoprudnyi,
Moscow Region, Russia}

\author{F. Pegoraro}
\affiliation{Department of Physics, University of Pisa  and CNISM, Pisa, Italy}

\date{November, 2005}

\begin{abstract}
The interaction of coherent nonlinear structures (such as
sub-cycle solitons, electron vortices and wake Langmuir waves)
with a strong wake wave in a collisionless plasma can be
exploited in order to produce ultra-short electromagnetic pulses.
The electromagnetic field of a coherent nonlinear structure is
partially reflected by the electron density modulations of the
incident wake wave and a single-cycle high-intensity electromagnetic
pulse is formed.
Due to the Doppler effect the length of this pulse is much
shorter than that of the coherent nonlinear structure.
This process is illustrated with two-dimensional
Particle-in-Cell simulations.
The considered laser-plasma interaction regimes
can be achieved in present day experiments
and can be used for plasma diagnostics.
\end{abstract}

\pacs{%
52.38.-r, 
42.65.Ky, 
52.35.Mw, 
52.35.Sb, 
52.35.We, 
47.32.Cc, 
52.70.-m, 
52.65.Rr} 


\maketitle

\section{Introduction}

Present day laser technology allows us to generate ultraintense laser pulses
with intensities approaching $10^{22}$ W/cm$^{2}$ \cite{Mourou}. The quiver
energy of the electrons in the electromagnetic (e.~m.) fields of such
intense pulses
is equal to, or greater than, their rest energy. This regime is achieved for
1$\mu$m wavelength pulses when the pulse intensity exceeds $10^{18}$ W/cm$%
^{2}$. In this regime the relativistic dynamics of the electrons
in the plasma inside which the pulse propagates, introduces a new
type of nonlinear phenomena (see e.~g., the review articles
\cite{Mourou} and \cite{SVB} and the literature quoted therein)
that arise from the nonlinearity of the Lorentz force and of the
relationship between particle momentum and velocity and, at very
large intensities, from nonlinear quantum electrodynamics effects
such as {electron-positron} pair creation \cite{SSb}.
It was soon realized that such
nonlinear processes can be harnessed (``relativistic
engineering'', as introduced in Ref. \cite{LightIntens} ) in order
to concentrate the e.~m. radiation in space and in time
and produce e.~m. pulses of unprecedented high intensity
or short duration that can be used to explore ultra-high energy
density effects in plasmas. The new possibilities that are made
available by nonlinear relativistic optics in plasmas were
emphasized by the results presented in \cite{Naumova}, where it is
shown that synchronized attosecond e.~m. pulses and
attosecond electron bunches can be produced during the interaction
of tightly focused, ultrashort laser pulses with overdense plasmas.
The property of nonlinear systems to respond anharmonically to a
periodic driving force was exploited in Ref. \cite{pipe} where the
propagation of a high intensity short laser pulse in a thin wall
hollow channel was shown to produce a coherent ultrashort pulse
with very short wavelength that propagates outwards through the
channel walls.

Recently, a different method of generating ultra-short e.~m.
pulses was proposed in Ref. \cite{IBKP}. This method uses the interaction
between a relativistic {electromagnetic sub-cycle} soliton and
the density modulations of a Langmuir wakefield in a plasma.
The mechanism envisaged is based on the results of
Ref. \cite{LightIntens}, where it was shown, that when a laser pulse
interacts with a breaking wake plasma wave, part of the pulse is reflected
in the form of a highly compressed and focused e.~m. pulse with an
up-shifted carrier frequency {due to the Doppler effect}.
The pulse  enhancement  of the pulse {intensity} and the
pulse compression arise because the electron density
modulations in the wake wave act as parabolic relativistic
mirrors. In the approach introduced in Ref.
\cite{IBKP} the role of laser pulse is taken by a
{sub-cycle} soliton
produced by another laser pulse in the plasma.

Relativistic {e.~m. sub-cycle} solitons are formed
during the interaction of a high intensity laser pulse with
a plasma and a significant
fraction of the pulse energy can be trapped in these structures,
up to of 30\% - 40\%, as was shown in Refs. \cite{Sol-Las-gen}.
This trapping occurs because, as
the laser pulse propagates in the plasma,
it loses part of its energy.
Since the number of photons in the pulse is approximately conserved,
the loss of pulse energy leads to the
down-shift of the pulse frequency
\cite{Sol-Las-gen,Sol-freq-down-2}
below the electron plasma frequency
(Langmuir frequency) $\omega_{pe} = (4\pi n_e e^2/m_e)^{1/2}$,
where $n_e$ is the electron density,
$e$ and $m_e$ -- electron charge and mass.
As a result, part of the pulse energy becomes trapped
inside electron density cavities in the form of low-frequency
radiation. The typical size of these solitons is of the order of
the collisionless electron skin depth $d_{e}=c/\omega _{pe}$. The
e.~m. fields inside the solitons consist of synchronously
oscillating electric and magnetic fields plus a steady
electrostatic field which arises from charge separation as
electrons are pushed outward by the ponderomotive force of the
oscillating fields, \cite{Sol-1,Sol-2}.
The development of the analytical
theory of intense e.~m. solitons
in collisionless plasma
can be traced in Refs.
\cite{Sol-Las-gen,Sol-freq-down-2,Sol-1,Sol-2,Sol-3,post}.
On a long time scale, the effects of the ion
motion become important and the ponderomotive force forms
quasi-neutral cavities in the plasma density, named post-solitons
in Ref. \cite{post}. Post-solitons were observed experimentally in
Ref. \cite{B-1} with the use of the proton imaging technique
\cite{B-2}. However, for simplicity, in the present paper we shall
consider conditions when the effects of the ion motion can be
neglected.

On the basis of one-dimensional (1D) analytical calculations,
in Ref. \cite{IBKP} it was shown that
the e.~m. field of the sub-cycle soliton
is partially reflected by the electron density modulations of the wake wave
and that
the frequency of the reflected pulse is up-shifted
by a factor $2\gamma_{ph}^2$ as compared to the soliton frequency
and
its intensity is proportional to $\gamma^3_{ph}$,
where $\gamma_{ph}$ is the Lorentz factor
corresponding to the phase velocity $v_{ph}$ of the wake wave.
It was thus proven that attosecond pulses can in principle be
generated by exploiting the soliton-wakefield interaction.

The aim of the  present paper is twofold:
first,
we confirm and qualify the analytical results of Ref. \cite{IBKP}
on the basis of two-dimensional (2D) particle-in-cell (PIC)
simulations of the interaction of a breaking wake plasma wave with
a sub-cycle soliton;
second,
we extend the analytical results of Ref. \cite{IBKP}
to other types of coherent nonlinear structures --
an electron vortex and  a wake field,
and these two cases are also illustrated with 2D PIC simulations.
The parameters of  the  simulations presented here are deliberately chosen
so as to reveal  unambiguously the effect of the reflection
and the frequency up-shift, thus providing
a proof-of-principle numerical experiment.
Nevertheless, one can see that these effects
should occur in present-day experiments
or can be deliberately realized with contemporary
terawatt laser systems.

    The paper is organized as follows.
The interaction of a wake wave with a soliton is
discussed in Sec. \ref{sec:II}.
In the first part of this section, \ref{sec:II:1D},
we briefly recall the
main properties of the reflected pulse, as obtained
analytically for a soliton in the 1D approximation.
The results of the 2D numerical simulations are
presented in the second part, \ref{sec:II:PIC}, together with the
comparison with the 1D analytic results.
In the  two following  sections \ref{sec:III} and \ref{sec:IV}
we present two other possible mechanisms of ultrashort
pulse generation which are due to the interaction of a breaking
wake plasma wave with either an electron vortex or with another
wake plasma wave oriented in the perpendicular
direction. Each of these sections is   subdivided into two parts,
the first contains the 1D analytical model and the second
-- the results of 2D PIC simulations.
Finally, in Sec. \ref{sec:V} the main results and
conclusions are listed.


\section{Reflection of the electromagnetic field of a soliton}
\label{sec:II}

\subsection{One-dimensional theory}
\label{sec:II:1D}

In this section we recall the 1D results on the form of the e.~m. pulse
reflected in the soliton-wake field interaction obtained analytically in
Ref. \cite{IBKP} .

When an intense short laser pulse interacts with a plasma, it induces a
wakefield \cite{Tajima} consisting of nonlinear Langmuir waves with a phase
velocity $v_{ph}=\beta_{ph}c$ equal to the group velocity of the laser
pulse. If the laser pulse propagates in a low density plasma, the latter is
close to the speed of light in vacuum. The nonlinearity of the strong
wakefield leads to the steepening of its profile and to the formation of
sharply localized maxima (spikes) in the electron density \cite{Akh-Pol}.
At wavebreak (see Ref. \cite{SVB} and references therein) the
electron density in
the spikes tends to infinity,
\begin{equation}  \label{ne}
n_{e}(X)=\frac{n_{i0}\beta_{ph}}{\beta _{ph}-\beta_e(X)},
\end{equation}
but remains integrable.
Here $X=x-v_{ph}t$,
$n_{i0}$ is the ion density
(equal to the unperturbed electron density $n_{e0}$), and
the ratio $\beta_e$ between the speed of the electrons and the speed of
light varies $-\beta _{ph}$ to $\beta _{ph}$.
Close to the wave breaking conditions, we can write
\begin{equation}  \label{ne1}
n_{e}(X)\simeq n_{e0}[1+\lambda _{p}\delta (X)]/2,
\end{equation}
where $\lambda_{p}$ is the plasma  wavelength in the wave-breaking regime,
$\delta(X)$ is the Dirac delta-function.
This density spike partially reflects a counterpropagating e.~m.
wave, as shown in Refs. \cite{SVB,LightIntens}.

In Ref. \cite{IBKP} the reflection of a 1D, circularly polarized
soliton was considered. The soliton was described by
the dimensionless vector potential $e \mathbf{A}/ (m_{e}c^{2}) =
A(x,t)~( \mathbf{e}_{y}+i \mathbf{e}_{z} )$
(see \cite{Sol-1}),
\begin{equation}  \label{A}
A(x,t)=\frac{2\varepsilon (\Omega_S) \,
     \cosh \left[ \varepsilon(\Omega_S)
     \omega_{pe}x/c\right] \, e^{i\,\Omega_S\,t}}
{\cosh ^{2}\left[\varepsilon (\Omega_S)\omega_{pe}x/c\right]
     -\varepsilon^{2}(\Omega_S)},
\end{equation}
where $\Omega_S < \omega _{pe}$ is the soliton frequency, and
$\varepsilon (\Omega_S)=\left( 1-\Omega_S^{2}/\omega_{pe}^{2}\right) ^{1/2}$.

The properties of the reflected pulse were
derived by performing a Lorentz transformation to the reference
frame where the wake plasma wave is at rest. In this frame the
reflection coefficient
\begin{equation}\label{coeff}
\rho (\omega ^{\prime })=-\frac{q}{q-i\omega ^{\prime }},
\end{equation}
where $q=2\omega _{pe}(2\gamma _{ph})^{1/2}$, derived in Ref.
\cite{LightIntens} was used for the frequency components obtained by
Fourier expanding the soliton amplitude. The form and amplitude of
the reflected pulse in the moving frame were then obtained by an
inverse Fourier transform followed by the inverse Lorentz
transformation of the vector potential.
The explicit form of the reflected pulse
is presented in Ref. \cite{IBKP}
by Eqs. (22) to (28) and by Figs. 2 and 3.

The main
conclusion of this analysis is that the amplitudes of the electric
and magnetic fields in the reflected pulse are increased by the
{factor}
$\gamma_{ph}^{3/2}$, i.~e., the pulse intensity is proportional to
$\gamma _{ph}^{3}$, while its frequency is up-shifted by
$2\gamma_{ph}^{2}$.
This scaling indicates that
in a tenuous plasma the frequency up-shift of the reflected pulse,
and its related compression, would  be so large that it could lead to
the generation of attosecond pulses.
The Lorentz factor $\gamma_{ph}$ of the wakefield
generated by a laser pulse in plasma is
of the order of
$\gamma_{ph}\approx\omega_d/\omega _{pe}$,
where $\omega_d$ is the frequency of the laser pulse
{(driver)} that generates
the wake plasma wave (see Ref.  \cite{Tajima}).
The frequency up-shift is
$2\gamma_{ph}^{2}\Omega_S \approx
2\gamma_{ph}^{2}\omega _{pe} \approx
2\gamma _{ph}\omega _{d}$.
Thus, for a $1\,\mu m$ wavelength laser pulse,
corresponding to the critical plasma density
$n_{cr} = m_e \omega_d^2/ 4\pi e^2
\approx 10^{21}cm^{-3}$,
the factor $2\gamma _{ph}$ { that would be required to generate}
an attosecond reflected pulse is of
order $10^{3}$, i.~e. the density of the plasma must be
{ order of} $4\times 10^{15}cm^{-3}$.
Denoting by $a_S$ the
dimensionless amplitude of the soliton (defined in terms of the
soliton frequency in the laboratory frame), the intensity of the
e.~m. field {in the soliton} is given by
$I_S \approx (a_S/\gamma_{ph})^{2} \times 10^{18}W/cm^{2}$.
Then, the intensity of the reflected
e.~m. pulse is $I_{ref} \approx \gamma _{ph}^{3} I_S=
a_S^{2}\gamma_{ph} \times 10^{18}W/cm^{2}$.
According to Ref. \cite{LightIntens} the paraboloidal
shape of the breaking wake plasma wave focuses the reflected e.~m.
pulse, further increasing its amplitude.
In the case of the soliton, the enhanced scaling of the
reflected  wave intensity,
$I_{ref} \approx \gamma_{ph}^{5} I_S$,
{which} leads to
$\approx a_S^{2} \gamma_{ph}^{3} \times 10^{18}W/cm^{2}.$

\subsection{Two-dimensional PIC simulation}
\label{sec:II:PIC}

In Ref. \cite{IBKP} the
generation of single cycle electromagnetic pulses
during the interaction of the wake Langmuir wave
with  a relativistic soliton (and in the present paper with other
nonlinear coherent structures)
is predicted with the help of a one-dimensional model.
In order to take into account the effects of multi-dimensional geometry
and strongly nonlinear plasma dynamics,
as well as the influence of kinetic effects,
we performed 2D PIC simulations
using the code REMP based on the PIC method
and ``density decomposition scheme'' \cite{ES}.

\begin{figure}
\includegraphics{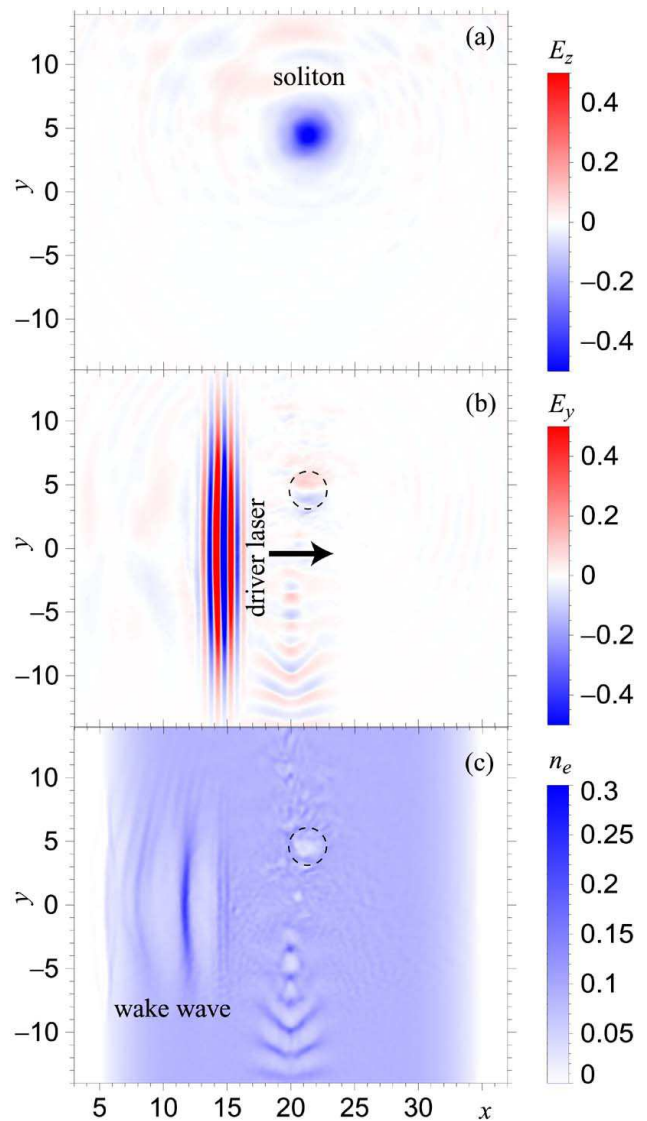}
\caption{\label{fig:S-1}
The electric field components $E_z$ (a) and $E_y$ (b)
and the electron density $n_e$ (c)
before the interaction between  the wake wave and the soliton,
$t = 61 \times 2\pi/\omega$.
The soliton is generated by the {auxiliary} laser pulse
which has already left the simulation box.
The dashed circle in (b), (c) denotes the soliton location.
}
\end{figure}

In the simulations presented here,
the grid mesh size is $\lambda_d/20$;
space and time unit is
$\lambda_d$ and $2\pi/\omega_d$, respectively.
Here $\lambda_d$ and $\omega_d$ is
the {driver} laser wavelength and frequency, respectively.
In the figures, the
electric and magnetic field components are normalized to
$m_e\omega_d c/e$
and the electron density is normalized to the critical density
$n_{cr} = m_e \omega_d^2 / 4\pi e^2$.

\begin{figure}
\includegraphics{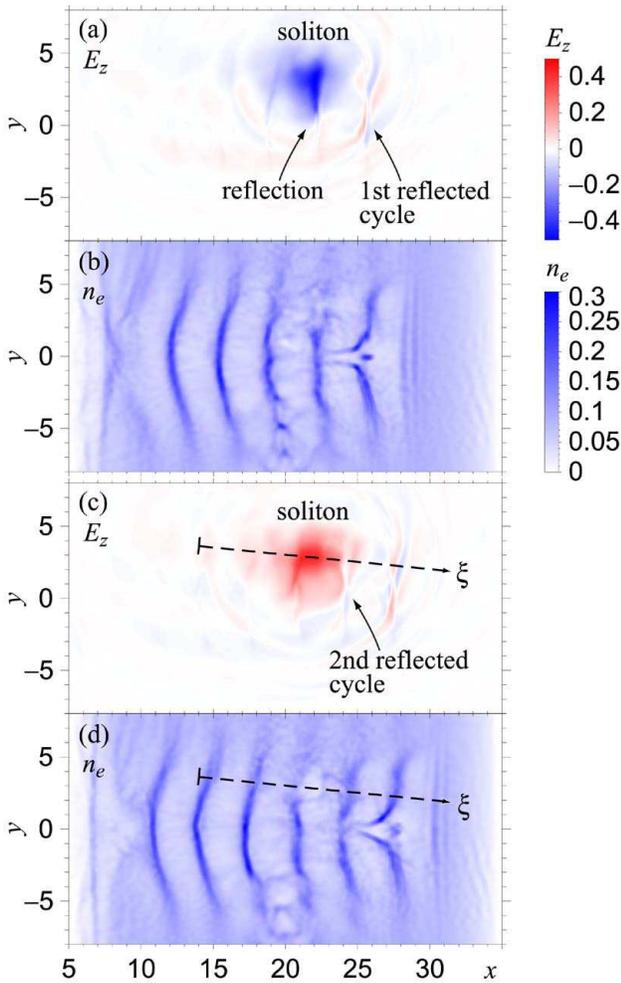}
\caption{\label{fig:S-2}
Interaction of the wake wave with the soliton.
The electric field component $E_z$ (a,c) and
electron density $n_e$ (b,d)
at $t = 76 $ (a,b)
and $78 \times 2\pi/\omega$ (c,d).
Dashed line $\xi$: see next figure.
}
\end{figure}

\begin{figure}
\includegraphics{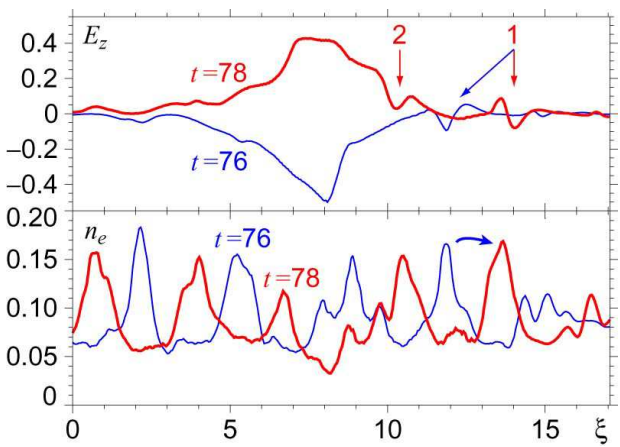}
\caption{\label{fig:S-3}
Reflection of the soliton electromagnetic field
by the electron density cusps.
The electric field component $E_z$ and
electron density $n_e$
at $t = 76, 78 \times 2\pi/\omega$
along the dashed line $\xi$ on the previous figure.
The first and second reflected pulses
are marked with ``1'' and ``2'', respectively.
}
\end{figure}

The ions are assumed to form an immobile neutralizing background
and thus only the electron motion is taken into account.
This approximation is applicable because
the typical interaction period is much shorter than the
ion response time, e.~g., in a hydrogen plasma.
In the simulations,
the boundary conditions are absorbing for the
e.~m. field and the quasi-particles.
The absorbing condition for the  e.~m. fields is implemented
using the scheme \cite{Tajima-Lee}
at the cost of an additional (absorbing) edge in the simulation box.

The interaction of a wake wave with a soliton is
simulated in a box with size $60\lambda_d \times 40\lambda_d$,
including the absorbing edges of thickness $3\lambda_d$.
The results are shown in Figs. \ref{fig:S-1}-\ref{fig:S-3}.
A single relativistic e.~m. sub-cycle soliton
is generated by an {auxiliary} laser pulse
with wavelength $\lambda_a = 2\lambda_d$ and
dimensionless amplitude $a_a = 0.5$,
corresponding to the peak intensity
$a_a^2 \times I_1$, where
$I_1 = 1.37\times 10^{18}\;$W/cm$^2\times(1\;\mu m/\lambda_d^2)$.
The pulse is Gaussian with FWHM size (length$\times$waist)
$4\lambda_d \times 6\lambda_d$.
The {auxiliary} laser pulse is linearly polarized
with its electric field along the $z$-axis;
it is generated at the bottom boundary
at $t=0$
and propagates along the $y$-axis at $x=20$.
The plasma wakefield, which interacts with the soliton,
is formed by a Gaussian  laser pulse, the {driver} pulse,
with amplitude $a_d =1.5$ and FWHM size $2\lambda_d \times 12\lambda_d$,
starting at time $t=45$ from the left boundary and
propagating along the $x$-axis.
The {driver} laser pulse is linearly polarized,
its electric field is directed along the $y$-axis.
The plasma slab occupies the region
$5 \le x\le 35$, $5 \le y\le 35$;
it is homogeneous in the direction of the $y$-axis
and it has convex parabolic slopes along the $x$-axis
from $x=5$ to $11$ and from $29$ to $35$.
This plasma-vacuum interface profile
is chosen so as to make the
laser pulse entrance into  the plasma smoother
and to avoid a fast wake wave breaking
that could happen in the case of a sharp plasma boundary.
The electron density at the center of the plasma slab
is $n_e = 0.09 n_{cr}$,
corresponding to the Langmuir frequency
$\omega_{pe} = 0.3$.
The number of quasi-particles is $3.24 \times 10^6$.

The phase velocity of the wakefield when it starts to interact with
the soliton is $v_{ph} \approx 0.925 $, corresponding to the Lorentz factor
$\gamma_{ph} \approx 2.63 $. 
The Lorentz factor $\gamma_{ph}$ is substantially smaller  than the  ratio
between the {driver} laser frequency and the
     plasma frequency because, in the case of short pulses,
the laser pulse group velocity strongly depends on the
pulse size \cite{group-vel-95}.

Fig. \ref{fig:S-1} shows
a portion  of the simulation box
shortly before the interaction.
The {auxiliary} laser pulse has already gone through:
in its wake we see
a single $s$-polarized relativistic e.~m. sub-cycle soliton
and remnants of a  broken wakefield at the bottom of the window.
The soliton frequency is well below the unperturbed plasma frequency,
$\Omega_S \approx 0.25 \omega_d < \omega_{pe}$.
The soliton appears as a region of low
electron density, as consistent with the fact that the electrons
are pushed outwards by the ponderomotive force of the oscillating
e.~m. fields inside the soliton.
Since the {driver} and the {auxiliary} laser pulses
have different polarizations and
the soliton inherits its polarization from the {auxiliary} laser,
it is easy to distinguish the e.~m. field reflected from the soliton
in the distribution of  the $E_z$ component.
The {driver} laser pulse induces a strong wakefield
which is seen in the electron density distribution
as a series of wide regions of rarefaction alternating
with   thin horseshoe-shaped regions of compression.
This is  the typical pattern of the wake of a Gaussian
intense short laser pulse.
Regions of compression correspond
to spikes (cusps) in the longitudinal profile of the electron density,
which is well explained by the one-dimensional theory \cite{Akh-Pol}.
These density cusps play the role of   semi-transparent mirrors,
moving with relativistic velocity.

Figs. \ref{fig:S-2}  and \ref{fig:S-3} show
the interaction of the density cusps
in the wake of the {driver} pulse with the soliton.
In Fig. \ref{fig:S-2}
the $z$-component of electric field
and the electron density are shown.
The wake wave of the {driver}
is close to the wave breaking regime.
Each electron density maximum (each cusp) in the wake acts as a fast moving
semitransparent parabolic mirror that partially reflects the e.~m.
fields of the soliton as it propagates through the soliton.
The process is repeated when the subsequent cusps of
the electron density propagate through the soliton.
Thus a set of short e.~m. pulses is formed.
We note that individual electrons perform an oscillatory motion,
while the electron density cusps exhibit a progressive motion.
According to Maxwell equations this electric charge density,
and the associated electric current density, determine
the e.~m. field evolution and reflection.
Even though the electron density cusp
is substantially distorted as it moves through the soliton,
it recovers after leaving the soliton.
Surprisingly, this transient distortion of the cusp
when crossing  the soliton
does not prevent the  formation of well pronounced single-cycle pulses,
and one can see the process of their generation even inside
the soliton.
We also note that the single cycle pulses move faster
than the electron ridge.

The frequency of the
fields in the reflected single-cycle e.~m. pulses is up-shifted and
their longitudinal size is much smaller than the size of the
soliton, as it can be clearly seen from Fig. \ref{fig:S-3}.
These simulation results confirm the physical mechanism investigated
analytically in Ref. \cite{IBKP}, as summarized in Sec. \ref{sec:II:1D}.
Since the soliton is not
exactly positioned at the crossing of laser pulse axes, the
reflected pulse is not exactly directed along the $x$-axis. This
is a consequence of the parabolic profile of the wakefield: as the
pulse is reflected by the upper wing of parabola it propagates at
an angle with respect to the $x$-axis.


Let us now estimate the parameters of the reflected
single-cycle pulse according to the results of the 1D analytic
calculations, using the initial conditions of the numerical
simulations. In this case $\gamma_{ph} \approx 2.63$.
The intensity of the
soliton is $I_S=(a_S \Omega_S/\omega_d)^2 I_1 \approx 0.016 I_1$,
where $a_S \approx 0.5$ is the soliton amplitude.
The reflected pulse intensity according to
the 1D analytic prediction is
$I_{ref}^{1D} = \gamma_{ph}^3 I_S \approx 0.28 I_1$.
The reflected pulse amplitude,
as seen from the results of numerical simulations,
is $a_{ref}\approx 0.14$.
Then $I_{ref} =
(2\gamma_{ph}^2 a_{ref}\Omega_S/\omega_d)^2 I_1 \approx 0.23 I_1$,
which is in reasonable agreement with the 1D theoretical analysis and
reproduce well the predicted scaling.


\section{Reflection of the electromagnetic field of the electron vortex}
\label{sec:III}

\subsection{One-dimensional theory}
\label{sec:III:1D}

Ultrashort electromagnetic pulses can also be generated in the
interaction of  a breaking wake plasma wave with different  types of
coherent structures besides solitons, such as electron vortices,
wake fields, etc. In this section we consider the interaction of a
wakefield with an electron vortex.
For the sake of illustration, in the following we will  employ a
simplified representation of their spatial structure.

\begin{figure}
\includegraphics[width=7cm]{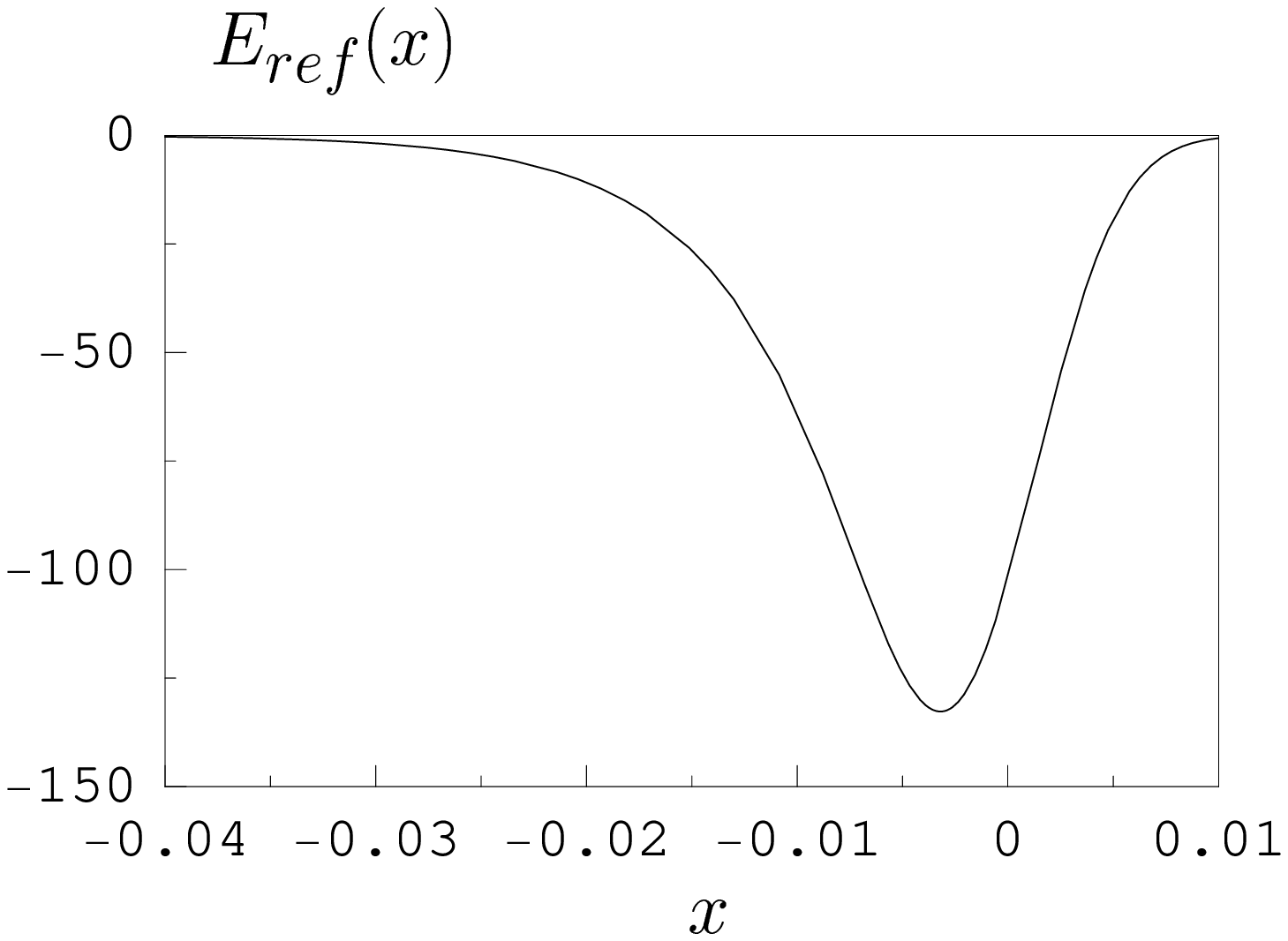}
\hspace{1.2cm}
\includegraphics[width=7cm]{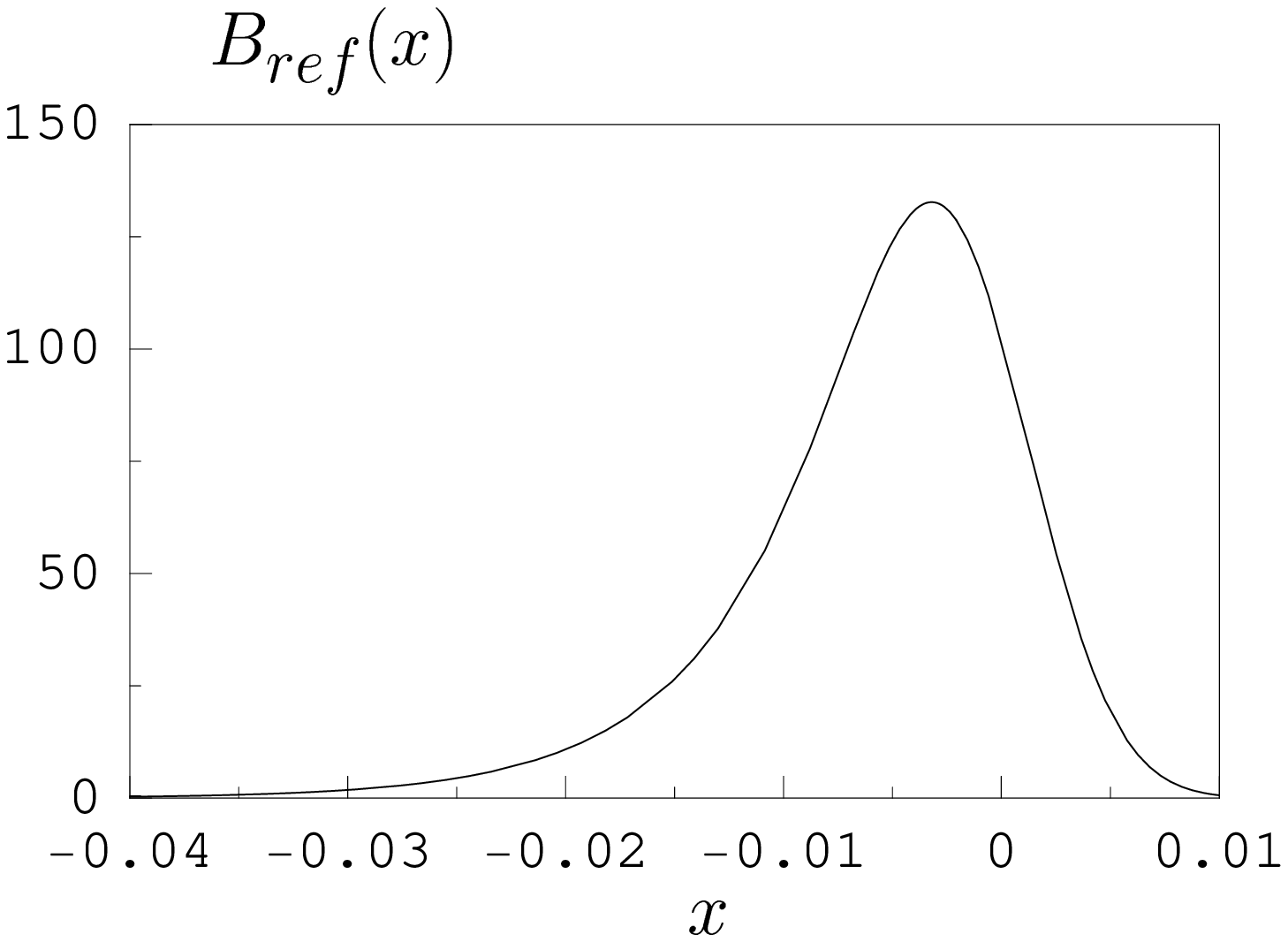}
\caption{\label{fig:WV-an}
The electric $E_{ref}$ and magnetic $B_{ref}$ fields of the reflected
pulse in the vortex-wakefield interaction as obtained from the
one-dimensional theory {\it vs.}
the spatial coordinate $x$ at
$t=0$, for  $\gamma_{ph}=10$. The fields are measured in units of the
initial vortex field  and $x$ is measured in  units of $c/\omega_{pe}$.}
\end{figure}

An electron vortex is characterized by a quasi-static magnetic field
generated by the current of  electrons circulating in a plasma with
steady ions.
Contrary to  hydrodynamic   vortices in ideal  fluids,
electron vortices in a plasma  have a characteristic
spatial scale that is given  approximately by
the  {plasma} collisionless skin depth $d_{e}=c/\omega_{pe}$.
The generation of  electron vortices,
together with their associated magnetic field,
    {by ultraintense laser pulses in plasmas is}
observed routinely in computer simulations
\cite{VORT} (see also \cite{SVB, VORT2} and  \cite{FINAL}).

For the sake of simplicity we consider a 1D configuration, which
corresponds to magnetic
``ribbons''  as discussed e.~g. in \cite{VORT,VORT2},   take the
magnetic field  to
be perpendicular to the direction of
the wake wave
and  assume  its profile  to be gaussian, i.~e.
\begin{equation}\label{b}
B=B_{0}\exp \left( -x^{2}/d_{e}^{2}\right).
\end{equation}
Following the procedure  proposed in Ref. \cite{IBKP}, we perform
the Lorentz transformation to the reference frame  where the
wakefield is at rest and the magnetic field of the electron vortex
appears as an incident e.~m. pulse. Then, we  Fourier
transform  the incident pulse and use the reflection and transition
coefficients  derived in Ref. \cite{LightIntens}. By inverting the
Fourier transform and by performing  the Lorentz transformation back
to the laboratory frame,  we obtain  that the shape of  the
reflected pulse is given by (see Fig. \ref{fig:WV-an})
\begin{equation} \label{vor}
\frac{E_{ref}}{\gamma_{ph} ^{3/2}B_{0}}=
\left( 8\pi \right)^{1/2}
\exp\left(\varphi^2-\psi\right)\left[1-
\mbox{erf}\left(\varphi-\frac{\psi}{2\varphi}\right)\right] ,
\end{equation}
where $\varphi = 2^{1/2}/\beta _{ph}\gamma_{ph} ^{1/2}$,
$\psi = 4\sqrt{2}\gamma_{ph} ^{3/2}\omega _{pe}(t-x/c)$,
and
$\mbox{erf}(x)= \frac{2}{\sqrt{\pi}}\int\limits_0^x\exp(-t^2)dt$
is the error function.
As expected, the field of the reflected pulse
scales for large $\gamma_{ph}$ as $\gamma_{ph} ^{3/2}$, and thus its
intensity {scales} as $\gamma_{ph} ^{3}$.
However the reflected pulse width
has a more complicated scaling. As can be clearly seen from  Fig.
\ref{fig:WV-an},
where the electric
and magnetic fields of the reflected pulse are shown, we have two
different scales
$1/\gamma_{ph} ^{2}$ and $1/\gamma_{ph} ^{3/2}$, and the front of the
reflected pulse is
much  steeper than its tail. This behaviour can also be deduced from
Eq. (\ref{vor}) since
the  width of the front of the pulse is determined by the expression
in square brackets
where the argument of the error function scales as $1/\gamma_{ph}
^{2}$. On the contrary,
the form of the tail is determined by the factor
$\exp\left(-\psi\right)$ where $\psi$ is
proportional to $\gamma_{ph} ^{3/2}$.

\subsection{Two-dimensional PIC simulation}
\label{sec:III:PIC}

\begin{figure}
\includegraphics{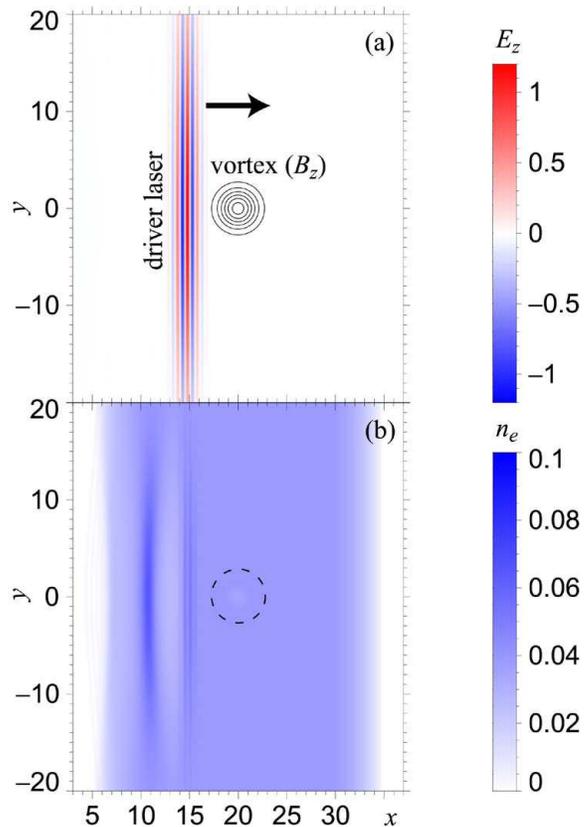}
\caption{\label{fig:V-1}
The electric field component $E_z$ (color-scale)
and contours of the magnetic field component $B_z$
for values $B_z = 1,2,3,4,5,6 \times 10^{-2}$
(a)
and the electron density $n_e$ (b)
at $t = 27 \times 2\pi/\omega$.
The dashed circle denotes the electron vortex location.
}
\end{figure}

\begin{figure}
\includegraphics{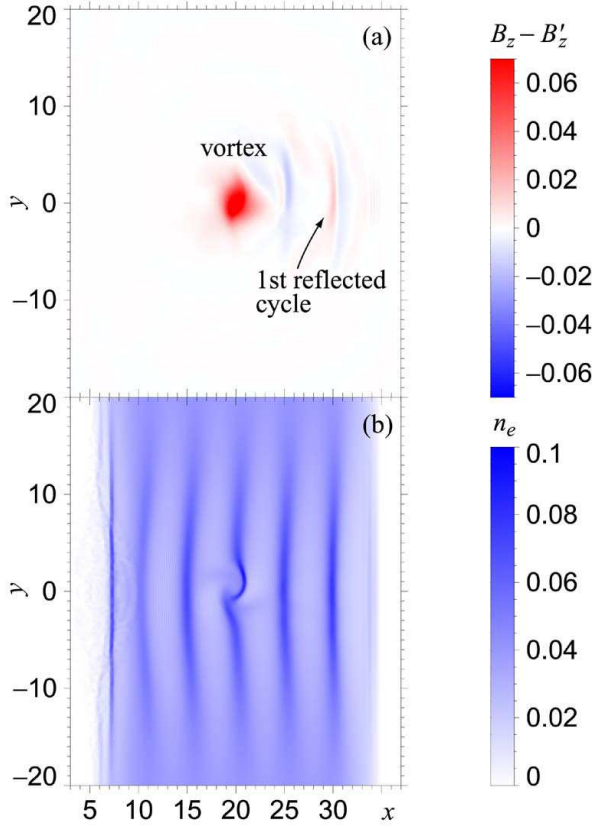}
\caption{\label{fig:V-2}
Interaction of the wake wave with the electron vortex,
$t = 47 \times 2\pi/\omega$.
(a) The difference between
the magnetic field component $B_z$
and $B'_z$, where  $B'_z$ is obtained from the simulation without the vortex.
(b) The electron density.
}
\end{figure}

\begin{figure}
\includegraphics{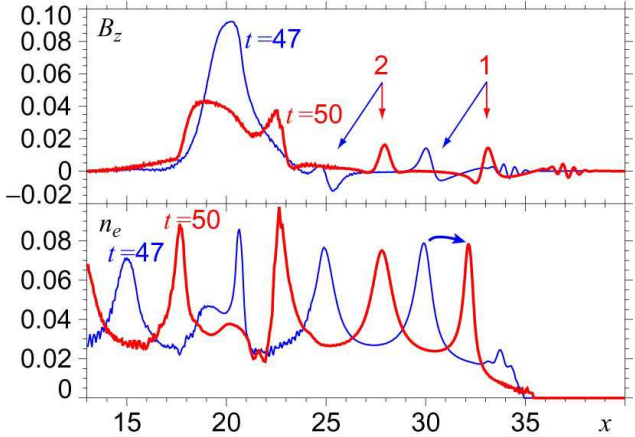}
\caption{\label{fig:V-3}
Reflection of the electromagnetic field of the electron vortex
by electron density cusps.
The magnetic field component $B_z$ and
electron density $n_e$ at $t = 47, 50 \times 2\pi/\omega$
along the $x$-axis at $y=0$.
The first and second reflected pulses
are marked with ``1'' and ``2'', respectively.
}
\end{figure}

The interaction of a wake wave with an electron vortex,
associated with a quasi-stationary transverse magnetic field,
is simulated in a box with size $50\lambda_d \times 80\lambda_d$,
including absorbing edges of thickness $2\lambda_d$.
The results are shown in Figs. \ref{fig:V-1}-\ref{fig:V-3}.
The electron vortex is prepared by introducing
a magnetic field $B_z$ which increases gradually
from $t=0$ to $t=10$.
After $t=10$, a self-consistent quasi-stationary electron fluid vortex
is formed with its corresponding   magnetic field distribution
$B_z = 0.066 \exp\big( -(x-20)^2/4 - y^2/4 \big)$,
Fig. \ref{fig:V-1}.
     The shape  of the plasma slab is the same as in the previous simulation
(Sec. \ref{sec:II:PIC})
except that its transverse size (along the $y$-axis) is $70\lambda_d$
and the maximum electron density is $n_e = 0.04 n_{cr}$.
The number of quasi-particles is $ 3.4 \times 10^6 $.
A Gaussian, linearly polarized ($E$ in the $z$-direction), {driver}
laser pulse with amplitude $a_d = 1.2$
and FWHM size $2\lambda_d \times 30\lambda_d$,
starts from $t=10$ from the left boundary, propagates along the $x$-axis,
and induces a plasma wakefield which  interacts with the electron vortex.
The phase velocity of the wakefield is $v_{ph} \approx 0.965 $,
so that the Lorentz factor is $\gamma_{ph} \approx 3.81 $.
A large waist of the {driver} laser pulse is chosen
to make the  interpretation of the reflection easier.

Fig. \ref{fig:V-2} shows the $z$-component of magnetic field
and the electron density during the interaction.
The wake wave itself has a transverse magnetic field $B'_z$
which arises due to the wake wave curvature.
The wake wave bending implies
that, in addition to the longitudinal oscillatory motion,
electrons move towards the wake wave axis, thus inducing
a weak transverse magnetic field which is maximum
on the periphery, in the regions of the electron density compression.
Eventually, this effect results in a so-called transverse wave break
of the wakefield \cite{TWB}.
Although the magnetic field of the wake wave
does not prevent us from   distinguishing   the
e.~m. pulses reflected from the vortex easily,
nevertheless, in order to improve the presentation,
in Fig. \ref{fig:V-2}
we show the quantity $B_z-B'_z$,
where $B_z$ and $B'_z$
are the magnetic field components
seen, respectively, in two simulations --
one, as described in this section,
and the other  performed with the same parameters
but without the vortex (i.~e., with no initial magnetic field).
Thus a ``pure'' reflection of the  e.~m. field of the vortex,
resulting in formation of single-cycle pulses,
can be seen.
In Fig. \ref{fig:V-3} the magnetic field component $B_z$ and
electron charge density $n_e$ are shown
along the $x$-axis at $y=0$, where the magnetic field of the wake wave
is exactly zero.

As in the case of the interaction of the wake wave and a soliton,
the electron density cusp
is substantially distorted as it moves through the vortex
and it recovers after passing the vortex, Figs. \ref{fig:V-2}, \ref{fig:V-3}.
As in the soliton case  this transient  distortion of the cusp
does not affect the shape of the resulting
single-cycle pulses.
We also note that the electron vortex is distorted
due to modulations of the electron density in the wake wave.
The size of the vortex and the magnitude of corresponding magnetic field
change so that the angular momentum and
vorticity of the electron fluid are preserved.


\section{Reflection of the electromagnetic field of the plasma wake wave}
\label{sec:IV}

\subsection{One-dimensional theory}
\label{sec:IV:1D}

Let us now consider the interaction of two
wake plasma waves
which are oriented perpendicularly  {with respect} to each other.
We assume that
the amplitude of one of the two waves (the first) is much smaller
than that of the other (the second) and thus neglect the action of
the first on the second wave.
In this scheme the first wake wave provides the electric field
which is (partially) reflected by the relativistic mirrors
represented by the electron density spikes of
the second wake wave.

\begin{figure}
\includegraphics[width=7cm]{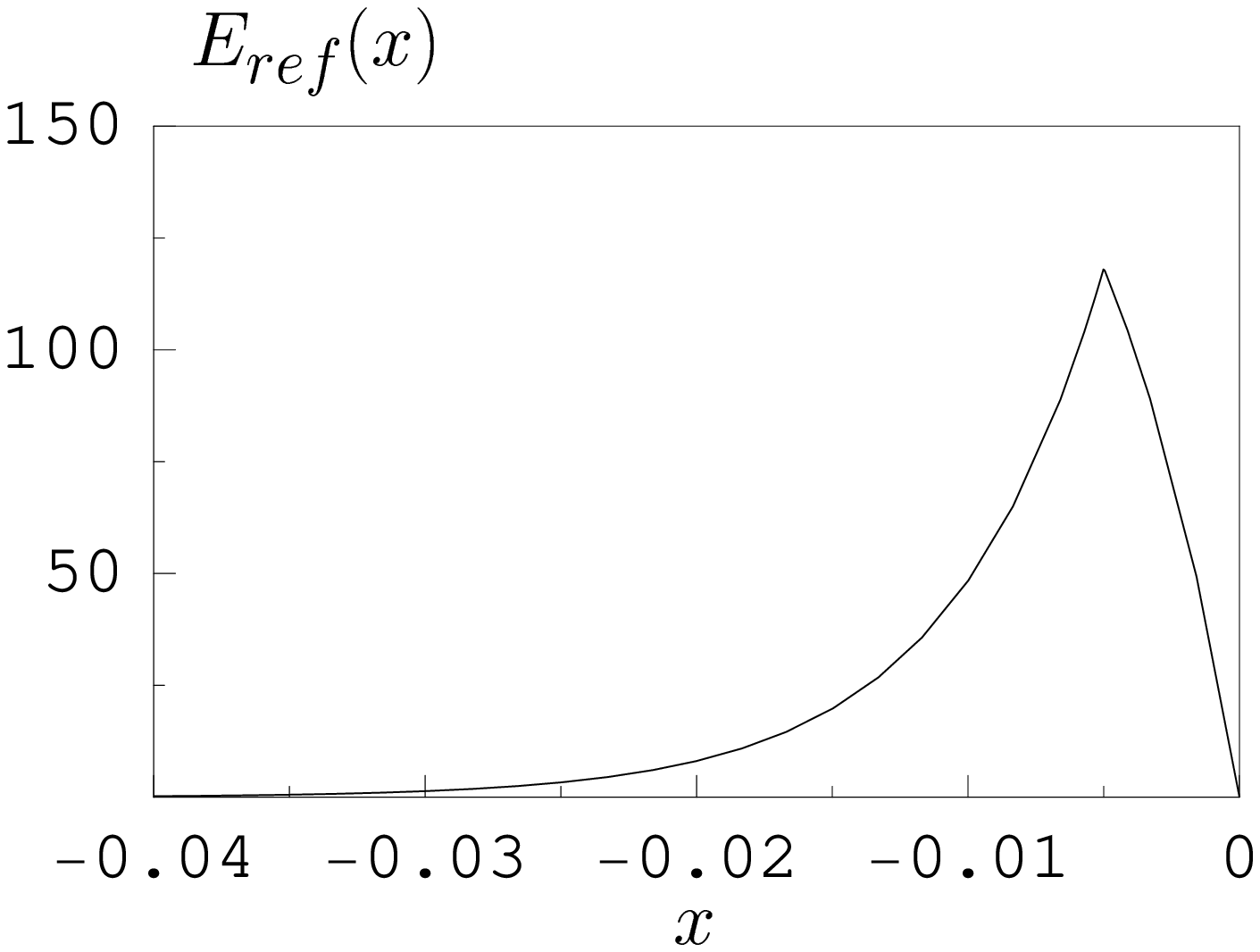}
\hspace{1.2cm}
\includegraphics[width=7cm]{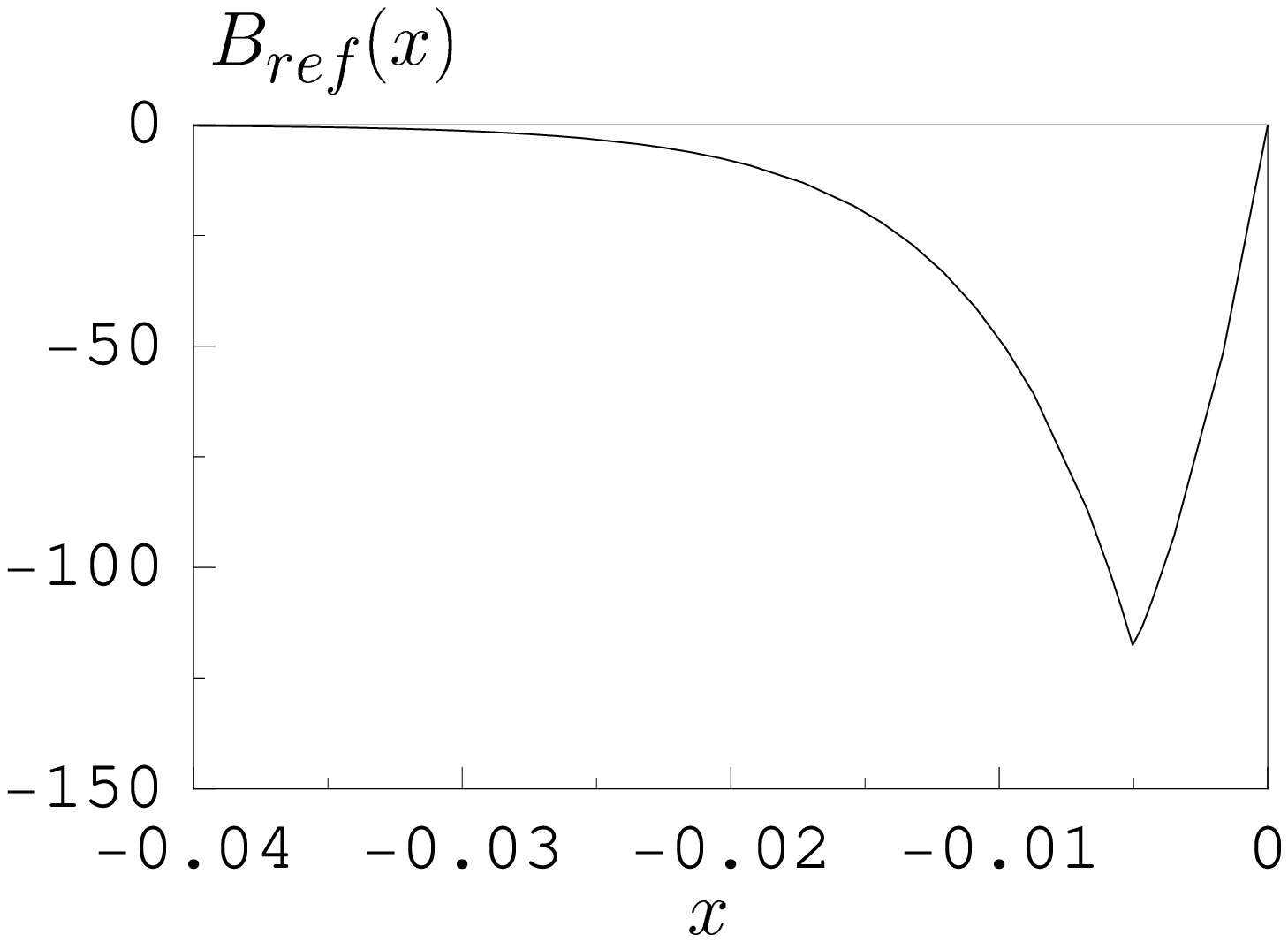}
\caption{\label{fig:WW-an}
The electric $E_{ref}$ and magnetic $B_{ref}$ fields of the reflected
pulse in the wakefield-wakefield interaction {\it vs.} the  spatial
coordinate $x$
at $t=0$ for  $\gamma_{ph}=10$. The fields are measured in  units of the
initial field strength of the first wakefield; $x$ is measured in
units of $c/\omega_{pe}$.}
\end{figure}

{We assume that}
the electric field of the first wake wave is directed along the
{$y$-axis}
and its spatial dependence  along the
$x$-axis, i.~e. along the propagation direction of
{the electron density spikes of} the second wave
(the mirrors),
can be modeled  as a two-step function with a constant
amplitude plateau in between.
We can choose the parameters of the
e.~m. pulse that generates the first wake wave  in such a
way that the  wavelength of the latter at   breaking is much larger
than its  transverse size (the size of the amplitude plateau) which
is assumed to be given by the characteristic scale of nonlinear
plasma structures $d_e=c/\omega_{pe}$, i.~e., such that
$2\sqrt{2\gamma_{ph}}c/\pi\omega_{pe}\gg d_e=c/\omega_{pe}$. In this
case  the electric field of the first wave during its reflection
from the second wake wave can be  taken  to be essentially a
function of $x$ only and to be of the form
\begin{equation} \label{wake}
E_w=E_0\theta(x)\theta(c/\omega_{pe}-x).
\end{equation}
Following the same procedure
{as in the  previous section},
we obtain for the electric field of
the reflected pulse
\begin{eqnarray} \label{wakeE}
E_{ref}=2\gamma_{ph}^2E_0\theta(t-x/c)\left[1-\exp(-\psi) \right.
\nonumber \\ \nonumber \\ \left.-
\left(1-\exp\left(\frac{2\sqrt{2}}{\sqrt{\gamma_{ph}}}-\psi\right)\right)
\theta\left((t-x/c)-\frac{1}{2\omega_{pe}\gamma_{ph}^2}\right)\right],
\end{eqnarray}
(see Fig. \ref{fig:WW-an}).
Here we also see the interplay of two scales: $1/\gamma_{ph}^2$ and
$1/\gamma_{ph}^{3/2}$ as in Fig. \ref{fig:WV-an}. The first one is
connected with
the width of the reflected pulse front and follows from the distance
between the arguments of the two theta functions. The second one determines the
tail width and arises from  the $\exp(-\psi)$ term in Eq. (\ref{wakeE}).
The reflected pulse field strength
reaches its maximum value
\begin{equation}
E_{max}=2\gamma^2_{ph}\left(1-\exp\left(\frac{2\sqrt{2}}
{\sqrt{\gamma_{ph}}}\right)\right)E_0
\end{equation}
at $t-x/c=1/(2\gamma^2_{ph}\omega_{pe})$.
For large values of
$\gamma_{ph}$ we obtain:
\begin{equation}
E_{max}=4\sqrt{2}\gamma^{3/2}_{ph}E_0.
\end{equation}
Thus the amplitude of the e.~m. field scales as
$\gamma^{3/2}_{ph}$, similarly to the case  of the reflection of the
soliton and of the electron vortex  from a breaking wake plasma
wave.
This scaling can be explained as follows.
First, the e.~m. field after reflection by
the moving mirror acquires a factor proportional to $\gamma^{2}_{ph}$
due to Doppler effect.
This can be easily shown with
the two Lorentz transformations
-- to the proper reference frame of the mirror
and back, to the laboratory reference frame
(the vector potential representing the e.~m. field
is perpendicular to the direction of propagation of the
mirror and thus remains unchanged while the frequencies and
wavenumbers are upshifted by a factor $\gamma^{2}_{ph}$).
Second, the reflection coefficient of the electron density spike
is proportional to $\gamma^{-1/2}_{ph}$.
These two facts result in the overall $\gamma^{3/2}_{ph}$ factor.
Actually, as discussed in the description of Figs. \ref{fig:WV-an}
and \ref{fig:WW-an}, the
spatial scaling of  the reflected pulse is not simply accounted for
by the frequency and wavenumber upshift.
In these two figures  we
observe two scales: $1/\gamma_{ph}^2$ and
$1/\gamma_{ph}^{3/2}$.
The first scale is a consequence of
the Doppler effect
and is connected with the frequency and
wavenumber upshift.
It manifests itself in the  frequency of the
reflected  {pulse from} a soliton and
in the scale of the front parts of the reflected
{pulses from} a vortex and a wakefield.
The second scale manifests itself,
in the case of the soliton, in
{the size of the reflected pulse envelope}
and, in the case of the vortex and of the wakefield,
in the decay length of {the reflected pulse tail}.
This second scale is
due to the preferential reflection of the low frequency part of the
field, i.~e., to the fact that the reflectivity of the mirror
increases and tends to unity for e.~m. radiation
with frequency of the order of
$\gamma_{ph}^{1/2}\omega_{pe}$ in the proper frame of the mirror
(or $\gamma^{3/2}_{ph}\omega_{pe}$, in the laboratory frame).

Finally, we should also note that  ultrashort e.~m. pulses
can be generated when a wake wave interacts with plasma in a
self-focusing channel \cite{self1}, see also, e.~g.,  \cite{self2}
and references therein.  In a self-focusing
channel,  an electric field is present due to  charge separation and
can be transformed
into an ultrashort e.~m. pulse,  similarly to the soliton,
vortex or wake wave.

\subsection{Two-dimensional PIC simulation}
\label{sec:IV:PIC}

\begin{figure}
\includegraphics[scale=0.9]{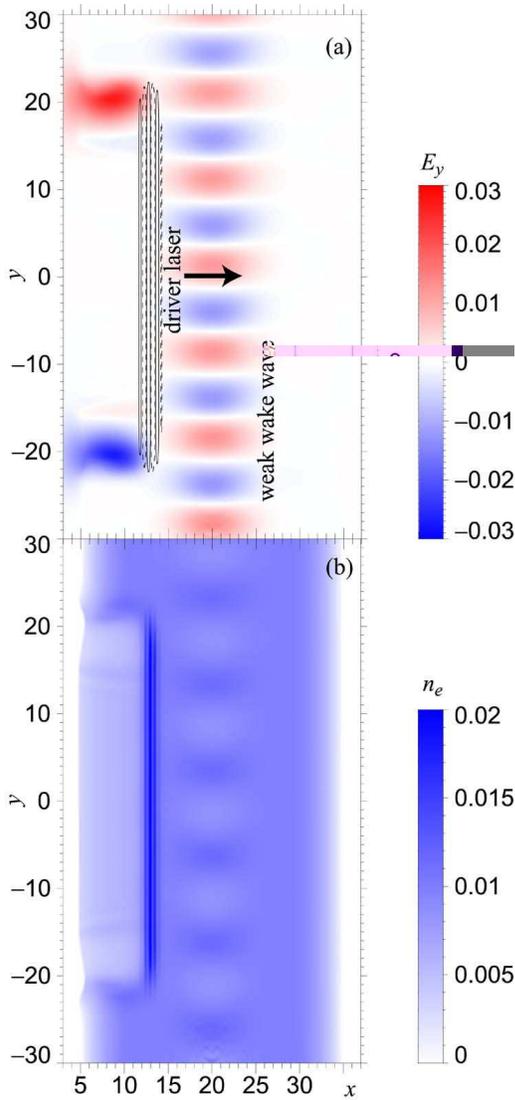}
\caption{\label{fig:W-1}
(a) The electric field component $E_y$ (color-scale)
and contours of the component $E_z$ for values
$E_z=-0.5$ (dashed) and $+0.5 m_e\omega c/e$ (solid)
and (b) the electron density $n_e$
at $t = 85 \times 2\pi/\omega$.
The weak wake wave, seen as a vertical periodic structure,
is generated by the {auxiliary} laser pulse
which has already left the simulation box.
In the wake from the {driver} laser pulse,
the $E_y$ is zero in the region $-10<y<10$,
which corresponds to the homogeneous part of the {driver}.
}
\end{figure}

\begin{figure}
\includegraphics[scale=0.9]{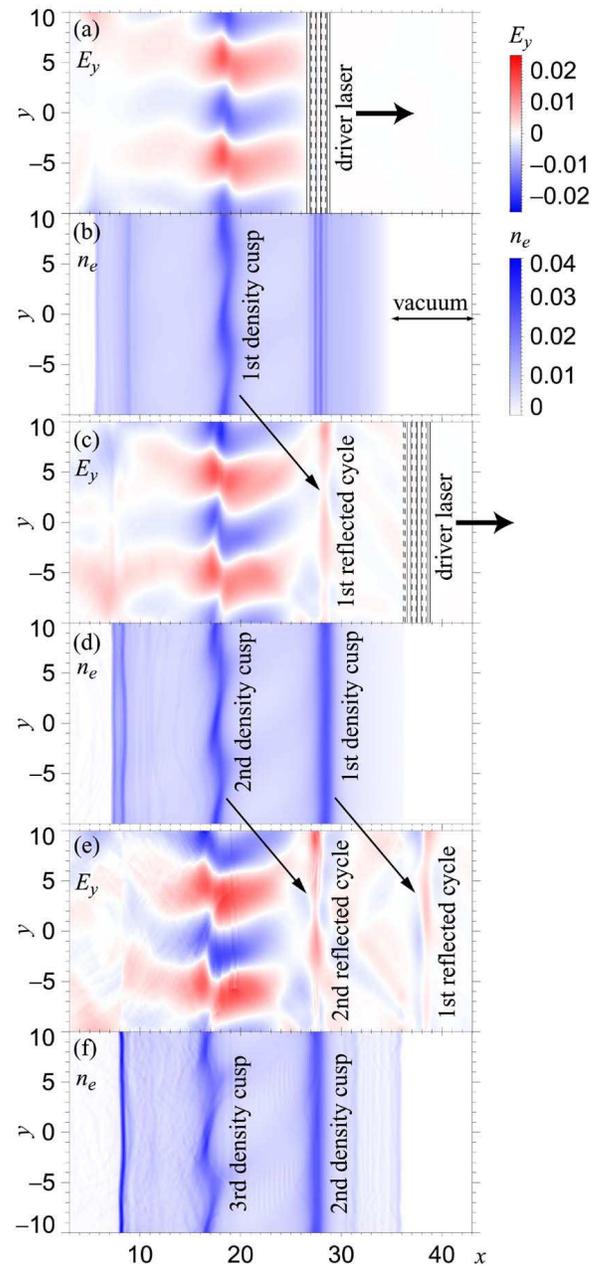}
\caption{\label{fig:W-2}
Interaction of the wake wave with the weak wave wave,
shown in the portion of the simulation box, $-10<y<10$,
corresponding to the quasi-one-dimensional part of
the wakefield from the {driver}.
The electric field component $E_y$ (a,c,e)
and electron density $n_e$ (b,d,f)
at $t = 100$ (a,b), $110$ (c,d) and $120 \times 2\pi/\omega$ (e,f).
The {driver} laser pulse is represented
by contours of the electric field component $E_z$
in the same way as in the previous figure.
}
\end{figure}

\begin{figure}
\includegraphics{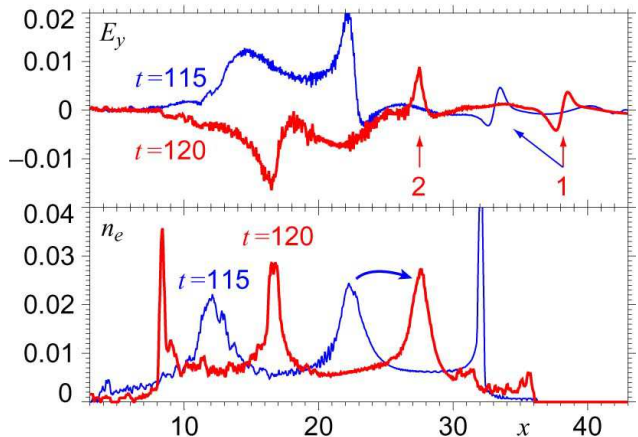}
\caption{\label{fig:W-3}
Reflection of the electromagnetic field of the weak wake wave
by electron density cusps of the wake wave from the {driver}.
The electric field component $E_y$ and
electron density $n_e$ at $t = 115, 120 \times 2\pi/\omega$
along the $x$-axis at $y=0$.
The first and second reflected pulses
are marked with ``1'' and ``2'', respectively.
}
\end{figure}

In order to show the interaction of a wake wave with another (weak) wake wave,
we performed a simulation
with the same simulation box and plasma distribution
as in the previous cases
(Sec. \ref{sec:III:PIC})
except for the maximum electron density
which is now $n_e = 0.01 n_{cr}$.
The results are shown in Figs. \ref{fig:W-1}-\ref{fig:W-3}.
Two laser pulses, which are linearly polarized
($E$ along the $z$-axis) and propagate in perpendicular directions,
induce two plasma wakes, Fig. \ref{fig:W-1}.
The first, {auxiliary}, laser pulse with Gaussian shape,
amplitude $a_a=0.7$, wavelength $\lambda_a = \lambda_d $
and FWHM size (length$\times$waist) $2.5\lambda_d \times 10\lambda_d$,
starts at $t=0$ from the bottom boundary,
propagates in the direction of the $y$-axis at $x=20$
and induces a weak wakefield with a longitudinal electric field
directed along the $y$-axis.
In Fig. \ref{fig:W-1} this
weak wake wave is seen as a vertical periodic structure.
The second laser pulse ({driver}) is switched on at $t=70$
on the left boundary.
Its amplitude is $a_d =2$ and size is $2\lambda_d \times 50\lambda_d$.
The {driver} pulse has a Gaussian profile
in the direction of the $x$ axis;
along the $y$-axis its shape is
a smooth centrosymmetric, piecewise polynomial
with sizes of adjacent parts $8,34,8$.
The central part is constant; the profile of the first slope
is defined by $(3-2\eta)\eta^2$, $\eta=y/8$.
We chose such the shape of the {driver} laser pulse
so as to ensure the transverse homogeneity
of the induced wakefield in the region $-17<y<17$,
and thus to obtain a plasma wake with zero
transverse electric field in this region.
In other words, the quasi-one-dimensional part
of the {driver} induces the wakefield
which is also quasi-one-dimensional.
With such a configuration,
in the distribution of the $E_y$ component inside the region $-17<y<17$,
we see, in principle, only the electric field of the
weak wakefield, induced by the {auxiliary} laser pulse.
The wakefield from the {driver} has phase velocity
$v_{ph} \approx 0.982 $ and the Lorentz factor
$\gamma_{ph} \approx 5.32 $.

Fig. \ref{fig:W-2} show
the $y$-component of electric field
and electron density at three moments of time
with period approximately equal to the period
of the plasma wake wave induced by the {driver} laser pulse.
The {driver} is represented by contours of the $z$-component
of electric field.
In Fig. \ref{fig:W-3} the electric field component $E_y$ and
electron charge density $n_e$ are shown
along the $x$-axis at $y=0$.

Again we note that, as in the case of the interaction of the wake wave
with a soliton and vortex,
the electron density cusp is distorted inside the weak wake wave,
but is almost restored when it moves outside, Figs. \ref{fig:W-2},
\ref{fig:W-3}.
Even though the electron density cusp seems curved
when it is under  the influence of the
weak wake wave, the single-cycle pulse reflected by this cusp
appears to be flat.

As in the cases of other nonlinear coherent structures
described above, each density cusp reflects
single-cycle e.~m. pulse whose frequency is up-shifted
and whose intensity is increased due to Doppler effect.
A complication, which is not included in the presented 1D theory,
arises from the fact that the weak wake wave,
which is (partially) reflected
by the ``mirrors'' of the wake from the {driver},
has almost the same phase velocity as these mirrors.


\section{Conclusion}
\label{sec:V}

Our analytical model and two-dimensional particle-in-cell simulations
show that during the interaction of coherent nonlinear structures (such as
sub-cycle solitons, electron vortices and wake Langmuir waves) with a
strong wake wave in a
collisionless plasma a train of single-cycle intense electromagnetic pulses
is generated. This effect
can be exploited in order to produce
ultra-short intense e.~m. pulses
with presently available lasers.
The results presented here confirm and extend the
analytical results obtained in Ref. \cite{IBKP}.

The modulations of electron density in the strong wake wave,
which is close to the wave-breaking regime,
have the shape  of spikes.
Each spike acts as a semi-transparent mirror
moving with a relativistic velocity,
corresponding to the phase velocity of the wake wave.
Such a mirror partially reflects
the electromagnetic field of a coherent nonlinear structure
and thus generates an electromagnetic pulse.
As predicted in Ref. \cite{IBKP},
the reflected pulse consists of a single cycle oscillation and,
as compared to the e.~m. field of the coherent nonlinear structure,
the reflected pulse has an up-shifted frequency and an increased intensity.
The reflected pulse intensity occurs as a result of
frequency up-shift, due to Doppler effect,
and because of the parabolic profile of the wake wave.

Using an analytical approach, we have shown that
in the three  cases  considered here of coherent nonlinear structures --
a sub-cycle soliton, an electron vortex and a wake wave --
the amplitude of the e.~m. pulse,
reflected by a relativistic flying mirror,
scales as $\gamma^{3/2}_{ph}$,
due to a similarity of the reflection process
in all three cases.
The reflection leads to a frequency and wavenumber upshift
which scales as $\gamma_{ph}^2$
and to the formation of an additional spatial scale
proportional to $1/\gamma_{ph}^{3/2}$.

Since in all the above cases of  wakefield interaction with nonlinear coherent
structures in  a plasma,   single-cycle e.~m. pulses are emitted
with a characteristic frequency, duration and polarization,
their emission represents  an important process to be used for
diagnostics of laser plasma interactions.

\section*{Acknowledgments}

The authors would like to acknowledge fruitful discussions with A.
Maksimchuk and V. Yanovsky. This work is partially supported by
INTAS Grant No. 001-0233, the Federal Program of the Russian
Ministry of Industry, Science and Technology N 40.052.1.1.1112. It
was also partially supported by RFBR Grant SS - 2328.2003.2.



\begin{thebibliography}{99}

\bibitem{Mourou}
G. A. Mourou, C. P. J. Barty, and M. D. Perry,
     {Phys. Today} \textbf{51}, 22 (1998).

\bibitem{SVB}
S. V. Bulanov, {et al.}, in \textit{Reviews of Plasma Physics},
edited by V. D. Shafranov
(Kluwer Academic/Plenum Publishers, New York, 2001), Vol. 22, p. 227.

\bibitem{SSb}
S. S. Bulanov, {Phys. Rev.} E \textbf{69}, 036408 (2004);
S. S. Bulanov, A. M. Fedotov, and F. Pegoraro,
     {Phys. Rev.} E \textbf{71}, 016404 (2005).

\bibitem{LightIntens}
S. V. Bulanov, T. Zh. Esirkepov, and T. Tajima,
     {Phys. Rev. Lett.} \textbf{91}, 085001 (2003).

\bibitem{Naumova}
N. M. Naumova, J. A. Nees, I. V. Sokolov, B. Hou, and G. A. Mourou,
     {Phys. Rev. Lett.} \textbf{92}, 063902 (2004);
N. Naumova, I. Sokolov, J. Nees, A. Maksimchuk, V. Yanovsky, and G. Mourou,
     {Phys. Rev. Lett.} \textbf{93}, 195003 (2004);
N. M. Naumova, J. A. Nees, and G. A. Mourou,
     {Phys. Plasmas} \textbf{12}, 056707 (2005).

\bibitem{pipe}
S. V. Bulanov, T. Zh. Esirkepov, N. M. Naumova, and I. V. Sokolov,
     {Phys. Rev.} E \textbf{67},(2003).


\bibitem{IBKP}
A. V. Isanin, S. S. Bulanov, F. F. Kamenets, and F. Pegoraro,
     {Phys. Lett.} \textbf{A 337}, 107 (2005).


\bibitem{Sol-Las-gen}
S. V. Bulanov, I. N. Inovenkov, V. I. Kirsanov, N. M. Naumova, and A.
S. Sakharov,
     {Phys. Fluids} B \textbf{4}, 1935 (1992);
S. V. Bulanov, T. Zh. Esirkepov, F. F. Kamenets, and N. M. Naumova,
     {Plasma Phys. Reports} \textbf{21}, 550 (1995);
S. V. Bulanov, T. Zh. Esirkepov, N. M. Naumova, F. Pegoraro,
     and V. A. Vshivkov, {Phys. Rev. Lett.} \textbf{82}, 3440 (1999);
T. Zh. Esirkepov, K. Nishihara, S. V. Bulanov, and F. Pegoraro,
     {Phys. Rev. Lett.} \textbf{89}, 275002 (2002).

\bibitem{Sol-freq-down-2}
K. Mima, M. S. Jovanovic, Y. Sentoku, Z.-M. Sheng,
     M. M. \v{S}kori\'c, and T. Sato, {Phys. Plasmas} \textbf{8},
2349 (2001);
Baiwen Li, S. Ishiguro, M. M. \v{S}kori\'c, Min Song, and T. Sato,
     {Phys. Plasmas} \textbf{12}, 103103 (2005).



\bibitem{Sol-1}
J. H. Marburger and R. F. Tooper, {Phys. Rev. Lett.} \textbf{35}, 1001 (1975);
C. S. Lai, {Phys. Rev. Lett.} \textbf{36}, 966 (1976);
T. Zh. Esirkepov, F. F. Kamenets, S. V. Bulanov, and N. M. Naumova,
     {JETP Lett.} \textbf{68}, 36 (1998).

\bibitem{Sol-2}
V. A. Kozlov, A. G. Litvak, and E. V. Suvorov,{Sov. Phys. JETP}
\textbf{49}, 75 (1979);
P. K. Kaw, A. Sen, and T. Katsouleas, {Phys. Rev. Lett.} \textbf{68},
3172 (1992);
D. Farina, M. Lontano, and S. V. Bulanov, {Phys. Rev.} E \textbf{62},
4146 (2000);
D. Farina and S. V. Bulanov, {Phys. Rev. Lett.} \textbf{86}, 5289 (2001);
S. Poornakala, A. Das, A. Sen, and P. K. Kaw, {Phys. Plasmas}
\textbf{9}, 1820 (2002);
S. Poornakala, A. Das, P. K. Kaw, A. Sen, Z. M. Sheng, Y. Sentoku, K.
Mima, and K. Nishikawa, {Phys. Plasmas} \textbf{9}, 3802 (2002);
M. Lontano, M. Passoni, and S. V. Bulanov, {Phys. Plasmas}
\textbf{10}, 639 (2003).

\bibitem{Sol-3}
J. I. Gerstein and N. Tzoar, {Phys. Rev. Lett. } \textbf{35}, 934 (1975);
N. L. Tsintsadze and D. D. Tskhakaya, {Sov. Phys. JETP} \textbf{45},
252 (1977);
P. K. Shukla, N. N. Rao, M. Y. Yu, and N. L. Tsintsadze, {Phys. Rep.}
\textbf{138}, 1 (1986);
H. H. Kuehl and C. Y. Zhang, {Phys. Rev.} E \textbf{48}, 1316 (1993);
Y. S. Dimant, R. N. Sudan, and O. B. Shiryaev, {Phys. Plasmas}
\textbf{4}, 1489 (1997).


\bibitem{post}
N. M. Naumova, S. V. Bulanov, T. Zh. Esirkepov, D. Farina, K.
Nishihara, F. Pegoraro, and A.S. Sakharov, {Phys. Rev. Lett.}
\textbf{87}, 185004 (2001).

\bibitem{B-1}
M. Borghesi, S. Bulanov, D. H. Campbell, R. J. Clarke, T. Zh.
Esirkepov, M. Galimberti, L. Gizzi, A. J. MacKinnon, N. M. Naumova,
F. Pegoraro, H. Ruhl, A. Schiavi, and O. Willi, {Phys. Rev. Lett.}
\textbf{88}, 135002 (2002).

\bibitem{B-2}
M. Borghesi, D. H. Campbell, A. Schiavi, M. G. Haines, O. Willi, A.
J. MacKinnon, P. Patel, L. A.Gizzi, M. Galimberti, R. J. Clarke, F.
Pegoraro, H. Ruhl, and S. Bulanov, {Phys. Plasmas} \textbf{9}, 2214
(2002).

\bibitem{Tajima}
T. Tajima and J. Dawson, {Phys. Rev. Lett.} \textbf{43}, 267 (1979).

\bibitem{Akh-Pol}
A. I. Akhiezer and R. V. Polovin, {Sov. Phys. JETP} \textbf{30}, 915 (1956).

\bibitem{ES}
T. Zh. Esirkepov, Comput. Phys. Comm. \textbf{135}, 144 (2001).

\bibitem{Tajima-Lee}
T. Tajima and Y. C. Lee, {J. Comput. Phys.} \textbf{42}, 406 (1981).

\bibitem{group-vel-95}
E. Esarey, P. Sprangle, M. Pilloff, and J. Krall,
{\it J. Opt. Soc. Am.} B {\bf 12}, 1695 (1995).


\bibitem{VORT}
S. V. Bulanov, T. Zh. Esirkepov, M. Lontano, F. Pegoraro, and A. M.
Pukhov, {Phys. Rev. Lett.} \textbf{76}, 3562 (1996).

\bibitem{VORT2}
S. V. Bulanov, T. Zh. Esirkepov, M. Lontano, and F. Pegoraro, {Plasma
Phys. Reports} \textbf{23}, 660 (1997).

\bibitem{FINAL}
N. M. Naumova, J. Koga, K. Nakajima, T. Tajima, T.Zh. Esirkepov, S.
V. Bulanov, and F. Pegoraro, {Phys. Plasmas} \textbf{8}, 4149 (2001).

\bibitem{TWB}
S. V. Bulanov, F. Pegoraro, A. M. Pukhov, and A. S. Sakharov, {Phys.
Rev. Lett.} {\bf 78}, 4205 (1997).

\bibitem{self1}
G. A. Askar'yan, Sov. Phys. JETP {\bf 15}, 8 (1962);
Sov. Phys. Uspekhi {\bf 16}, 680 (1973);
A. G. Litvak, Sov. Phys. JETP {\bf 30}, 344 (1969);
C. Max, J. Arons, and A. B. Langdon, Phys. Rev. Lett. {\bf 33}, 209 (1974).

\bibitem{self2}
N. M. Naumova,  S. V. Bulanov,  K. Nishihara, T.Zh. Esirkepov, and F.
Pegoraro, {Phys. Rev.} E \textbf{  65}, 045402 (2002).

\end{thebibliography}
\end{document}